\documentclass[a4paper,12pt]{article}
\usepackage{amsmath,amsfonts}
\def\ba {\begin {array}}
\def\ea{\end {array}}
\def\be {\begin {equation}}
\def\ee {\end {equation}}
\def\bea {\begin {eqnarray}}
\def\eea {\end {eqnarray}}

\begin{document}
\title
{Superintegrability and higher order constants for  quantum systems}
\author{E. G.~Kalnins\\
Department of Mathematics,\\
University
of Waikato, Hamilton, New Zealand.\\
J.~M.~Kress\\School of Mathematics and Statistics,\\
 University of New South Wales,\\ Sydney, Australia.\\
W.~Miller, Jr.\\
 School of Mathematics, University of Minnesota,\\
Minneapolis, Minnesota, U.S.A.}

\maketitle

\begin{abstract} We refine a method for finding a canonical form of symmetry operators of arbitrary order for the Schr\"odinger eigenvalue equation $H\Psi \equiv (\Delta_2 +V)\Psi=E\Psi$ on any 2D Riemannian manifold, real or complex that admits a separation of variables in some orthogonal coordinate system. The flat space equations with potentials $
V=\alpha (x+iy)^{k-1}/{ (x-iy)^{k+1}}$ in Cartesian coordinates, and 
$V=\alpha r^2+ 
{\beta}/{ r^2\cos ^2k\theta } +{\gamma}/{ r^2\sin ^2k\theta }$ (the Tremblay, Turbiner, and Winternitz system) in polar coordinates,  have each been shown to be classically superintegrable for all rational numbers $k$. We apply the canonical operator method to give a constructive proof that each of these  systems is also quantum superintegrable.  We also develop the classical analog of the quantum canonical form for a symmetry. It is clear that our methods will generalize to other Hamiltonian systems.
\end{abstract}

\section{Introduction}  
We consider an $n$-dimensional classical superintegrable system is an integrable Hamiltonian 
system that not only possesses $n$ mutually commuting integrals, but in addition,
the Hamiltonian Poisson-commutes with $2n-1$ functions on the phase space that are globally
defined and polynomial in the momenta.  This notion can be extended to define a quantum
superintegrable system to be a quantum Hamiltonian which is one of a set
of $n$ independent mutually commuting differential operators that commutes with a set of $2n-1$
independent differential operators of finite order. We restrict to classical systems of the form ${\cal H}=\sum_{i,j=1}^ng^{ij}p_ip_j+V$ and corresponding quantum systems $H=\Delta_n+{\tilde V}$ where $\tilde V$ is related to $V$ but in general is not equal to it, \cite{Gravel, KKM20061}. These systems, including the classical Kepler problem and the quantum hydrogen atom have great historical importance, due to their remarkable properties. They are exactly analytically solvable and in multiple ways. Superintegrable systems of 1st order, i.e., classical systems where the defining symmetries are first order in the momenta and quantum systems where the symmetries are first order partial differential operators, are directly related to Lie transformation groups and well understood. Superintegrable systems  of 2nd order have been well studied and there is now a structure and classification theory \cite{ KKM20061,KKMP, DASK2005, KKM2007}. The connection between 2nd order symmetries and separation of variables has been of crucial importance in finding examples and carrying out the classification \cite{FSUW, MSVW,KKM2}. However for 3rd and higher order superintegrable systems much less is known. In particular there have been  relatively few examples and there  is almost no structure theory, i.e., an understanding of the structure of the Poisson  algebra generated by the classical symmetries or the algebra generated by the quantum symmetries and their commutators, and no classification theory.

This situation has changed recently with the discovery of many more examples of classical (especially) and quantum superintegrable systems of order higher than two, \cite{VE2008, Evans2008a, TTW, TTW2,KMP10, KKM10}. Also, the tool of coupling constant metamorphosis (St\"ackel transform) has been developed to map superintegrable systems of higher order on one Riemannian space to superintegrable systems of the same order and structure on a different Riemannian space \cite{HGDR,KR00, SB2008, ASB, JH}. In this paper we will concentrate on the case $n=2$ where the number of independent symmetries is $3$, including the classical Hamiltonian or quantum Schr\"odinger operator, respectively. In almost all of the families of new examples the second symmetry is of 2nd order and defines an orthogonal separable coordinate system for the classical Hamilton-Jacobi equation or the quantum Schr\"odinger equation. Only one defining symmetry is of higher order. We are particularly considering the classical example of Tremblay, Turbiner, and Winternitz \cite{TTW, TTW2} where 
$$V=\alpha r^2+ 
\frac{\beta}{ r^2\cos ^2k\theta } +\frac{\gamma}{ r^2\sin ^2k\theta }$$
in polar coordinates. Due to the separation in polar coordinates there are two commuting 2nd order constants of the motion. For certain rational values of the parameter $k$ these authors found an additional constant of the motion (usually of higher order), so that the system was superintegrable  both in the classical and quantum sense. They conjectured and provided impressive evidence that these systems were classically and quantum superintegrable for all rational $k$. In \cite{KMPog10} it was shown that, in fact all of the classical TTW systems were superintegrable. Quesne \cite{CQ10} used a structure developed by Dunkl to show that the TTW systems for $k$ an odd integer were quantum superintegrable. As a bi-product of the tools developed in this paper we will give a constructive proof  that the TTW system is quantum superintegrable for all rational $k$. However, our main contribution is a tool for the verification of classical and quantum superintegrability of higher order that can be applied to a variety of Hamiltonian systems.

In Section 1  we review a construction of a canonical form for symmetry operators of all orders of  a  time-independent
Schr\"odinger equation that admits an orthogonal separation of variables \cite{KMPog04, KKMPog07}. Then in Section 2 we apply this tool to the flat space potential   $
V=\alpha{(x+iy)^{k-1}}/{ (x-iy)^{k+1}}$ in Cartesian coordinates (separable in polar coordinates), that has recently been shown to be classically superintegrable for all rational $k$ \cite{KMPog10}. We demonstrate that it is also quantum superintegrable for all rational $k$. 

In Section 3 we give the analogous construction of a canonical form for symmetry operators of all orders of for a classical Hamiltonian system.  We again treat the example $
V=\alpha {(x+iy)^{k-1}}/{ (x-iy)^{k+1}}$ and give a new proof that it is classically superintegrable for all rational $k$. The special interest here is the relation between the classical and quantum construction.

In Section 4 we apply canonical operator method to the TTW case to demonstrate that it is quantum superintegrable for all rational $k$. In Section 5 we discuss our overall strategy and the prospects for exploitation and generalization of our methods.

\section{The canonical form for a symmetry operator}
We consider a Schr\"odinger equation on a 2D real or complex Riemannian manifold with Laplace-Beltrami operator $\Delta_2$ and potential $V$:
\be \label{TIS}H\Psi\equiv (\Delta_2+V)\Psi=E,\Psi \ee
that also  admits an orthogonal  separation of variables. 
If $\{u_1,u_2\}$ is the  orthogonal separable coordinate system   the corresponding Schr\"odinger
operator has the form 
\begin{equation}
H= L_1= \Delta_2+V(u_1,u_2)=
\frac{1}{f_1(u_1)+f_2(u_2)}\left(\partial^2_{u_1}+\partial^2_{u_2}+v_1(u_1)+v_2(u_2)
\right).\nonumber
\end{equation}
and, due to the separability, there is the second-order symmetry operator
\[
L_2= \frac{f_2(u_2)}{f_1(u_1)+f_2(u_2)}\left(\partial^2_{u_1}+v_1(u_1)\right)-\frac{f_1(u_1)}{f_1(u_1)+f_2(u_2)}\left(\partial^2_{u_2}+v_2(u_2)\right),
\]
i.e., $
[L_2,H]=0,$ 
and the operator identities 
\begin{equation} \label{fundident}
f_1(u_1)H+L_2=\partial^2_{u_1}+v_1(u_1),\qquad
f_2(u_2)H-L_2=\partial^2_{u_2}+v_2(u_2).
\end{equation}

We  look for a  partial differential  operator  ${\tilde L}(H,L_2,u_1,u_2)$ that satisfies 
\begin{equation}
[H,{\tilde L}]= 0.\label{newconditions1}
\end{equation}
We require that the symmetry operator take the standard form
\begin{equation}\label{standardLform}
{\tilde
L}=\sum_{j,k}\left(A^{j,k}(u_1,u_2)\partial_{u_1u_2}+B^{j,k}(u_1,u_2)\partial_{u_1}
+C^{j,k}(u_1,u_2)\partial_{u_2}+ D^{j,k}(u_1,u_2)
\right)H^jL_2^k.
\end{equation}
Note that if the formal operators  (\ref{standardLform})
contained partial
derivatives in $u_1$ and $u_2$ of orders $\ge 2$ we could use the  identities (\ref{fundident}), recursively, and 
 rearrange terms to achieve the
unique standard form (\ref{standardLform}).

Using operator identities and writing $\partial_{u_j}=\partial_j$ we have 
\[
[\partial_1,H]= -\frac{f'_1}{f_1+f_2}H+\frac{v'_1}{f_1+f_2},\quad
[\partial_2, H] = -\frac{f'_2}{f_1+f_2}H+\frac{v'_2}{f_1+f_2},
\]
\[
[\partial_1, L_2] = -\frac{f'_1f_2}{f_1+f_2}H+\frac{f_2v'_1}{f_1+f_2},\quad
[\partial_2, L_2] = \frac{f_1f'_2}{f_1+f_2}H-\frac{f_1v'_2}{f_1+f_2},
\]

\[
[H,\partial_{12}]=\frac{f_2'}{f_1+f_2}\partial_1 H+\frac{f_1'}{f_1+f_2}\partial_2 H-\frac{1}{f_1+f_2}(v_2'\partial_1+v_1'\partial_2),\]
\[ [H,F(u_1,u_2)]=\frac{1}{f_1+f_2}(F_{u_1u_1}+F_{u_2u_2}+2F_{u_1}\partial_{1}+2F_{u_2}\partial_{2}).\]
From these results and (\ref{fundident}) we obtain
\[(f_1+f_2) [H,A\partial_{12}]=2A_{u_1}(f_1\partial_2H+\partial_2 L_2-v_1\partial_2)+2A_{u_2}(f_2\partial_1 H-\partial_1 L_2-v_2\partial_1)\]
\[ +A(f_2\partial_1H+f_1'\partial_2 H -v_2'\partial_1-v_1'\partial_2)+(A_{u_1u_1}+A_{u_2u_2})\partial_{12},\]
\[(f_1+f_2) [H,B\partial_1]=B(f_1'H-v_1')+2B_{u_1}(f_1H+L_2-v_1)+(B_{u_1u_1}+B_{u_2u_2})\partial_1+2B_{u_2}\partial_{12},\]
\[(f_1+f_2) [H,C\partial_2]=C(f_2'H-v_2')+2C_{u_2}(f_2H-L_2-v_2)+(C_{u_1u_1}+C_{u_2u_2})\partial_2+2C_{u_1}\partial_{12},\]
\[(f_1+f_2) [H,D]=D_{u_1u_1}+D_{u_2u_2}+2D_{u_1}\partial_{1}+2D_{u_2}\partial_{2}.\]
Thus we have 
\[
(f_1(u_1)+f_2(u_2))[H, A(u_1,u_2)\partial_{12}+B(u_1,u_2)\partial_{1}+C(u_1,u_2)\partial_{2}+ D(u_1,u_2)]=
\]
\[
(A_{u_1u_1}+A_{u_2u_2}+2B_{u_2}+2C_{u_1})\partial_{12}+(B_{u_1u_1}+B_{u_2u_2}-2A_{u_2}v_2+2D_{u_1}-Av'_2)\partial_{1}
\]
\[
+(2A_{u_2}f_2+Af'_2)\partial_{1}H-2A_{u_2}\partial_{1}L_2+(C_{u_1u_1}+C_{u_2u_2}-2A_{u_1}v_1+2D_{u_2}-Av'_1)\partial_{2}
\]
\[
+(2A_{u_1}f_1+Af'_1)\partial_{2}H+2A_{u_1}\partial_{2}L_2
\]
\[+(D_{u_1u_1}+D_{u_2u_2}-2B_{u_1}v_1-2C_{u_2}v_2-Bv'_1-Cv'_2)
\]
\[
+(2B_{u_1}f_1+2C_{u_2}f_2+Bf'_1+Cf'_2)H+(2B_{u_1}-2C_{u_2})L_2.
\]

The symmetry condition (\ref{newconditions1}) is equivalent to the
system of equations
\begin{equation} \label{partialxy}
\partial_{11}A^{j,k}+\partial_{22}A^{j,k}+2\partial_{2}B^{j,k}+2\partial_{1}C^{j,k}
=0,
\end{equation}
\begin{eqnarray} \label{partialx}
\partial_{11}B^{j,k}+\partial_{22}B^{j,k}-2\partial_{2}A^{j,k}v_2+2\partial_{1}D^{j,k}-A^{j,k}v'_2+\nonumber\\
(2\partial_{2}A^{j-1,k}f_2+A^{j-1,k}f'_2)-2\partial_{2}A^{j,k-1}=0,&&
\end{eqnarray}
\begin{eqnarray} \label{partialy}
\partial_{11}C^{j,k}+\partial_{22}C^{j,k}-2\partial_{1}A^{j,k}v_1+2\partial_{2}D^{j,k}-A^{j,k}v'_1+&&\nonumber\\
(2\partial_{1}A^{j-1,k}f_1+A^{j-1,k}f'_1)+2\partial_{1}A^{j,k-1}=0,&&
\end{eqnarray}
\begin{equation}\label{constantterm}
\partial_{11}D^{j,k}+\partial_{22}D^{j,k}-2\partial_{1}B^{j,k}v_1-2\partial_{2}C^{j,k}v_2-B^{j,k}v'_1-C^{j,k}v'_2
\end{equation}
\[
+(2\partial_{1}B^{j-1,k}f_1+2\partial_{2}C^{j-1,k}f_2+B^{j-1,k}f'_1
+C^{j-1,k}f'_2)
+(2\partial_{1}B^{j,k-1}-2\partial_{2}C^{j,k-1})=0.
\]

Note that condition (\ref{standardLform}) makes sense, at least formally, for infinite order differential equations. Indeed, one can 
consider $H,L_2$ as parameters in these equations. Then once $\tilde L$ is
expanded as a power series in these parameters, the terms are reordered so that the powers of the parameters are on the right, before
they are replaced by explicit differential operators. Alternatively one can consider the operator $\tilde L$ as acting on a
simultaneous eigenbasis of the commuting operators $H$ and $L_2$, in
which case the parameters are the eigenvalues. Of course (\ref{standardLform}) is defined rigorously for finite order symmetry operators.

In this view we can write
\begin{eqnarray}\label{generalLform}
{\hat
L}(H,L_2,u_1,u_2)&=&A(u_1,u_2,H,L_2)\partial_{12}+B(u_1,u_2,H,L_2)\partial_{1}\nonumber\\
&+&C(u_1,u_2,H,L_2)\partial_{2}
+ D(u_1,u_2,H,L_2),
\end{eqnarray}
and consider $\hat  L$ as an at most second-order  order differential operator in $u_1,u_2$ that is analytic in the parameters $H,L_2$. 
Then the above system of equations can be written in the more compact form
\begin{equation} \label{partialxyHL}
A_{u_1u_1}+A_{u_2u_2}+2B_{u_2}+2C_{u_1}
=0,
\end{equation}
\begin{equation} \label{partialxHL}
B_{u_1u_1}+B_{u_2u_2}-2A_{u_2}v_2+2D_{u_1}-Av'_2+(2A_{u_2}f_2+Af'_2)H-2A_{u_2}L_2=0,
\end{equation}
\begin{equation} \label{partialyHL}
C_{u_1u_1}+C_{u_2u_2}-2A_{u_1}v_1+2D_{u_2}-Av'_1+(2A_{u_1}f_1+Af'_1)H+2A_{u_1}L_2=0,
\end{equation}
\begin{equation}\label{constanttermHL}
D_{u_1u_1}+D_{u_2u_2}-2B_{u_1}v_1-2C_{u_2}v_2-Bv'_1-Cv'_2
\end{equation}
\[
+(2B_{u_1}f_1+2C_{u_2}f_2+Bf'_1
+Cf'_2)H+(2B_{u_1}-2C_{u_2})L_2=0.
\]
We can view (\ref{partialxyHL}) as an equation for $A,B,C$ and (\ref{partialxHL}), (\ref{partialyHL}) as the defining equations for $D_{u_1}, D_{u_2}$.
Then $\tilde L$ is $\hat L$ with the terms in $H$ and $L_2$ interpeted as (\ref{standardLform}) and considered as partial differential operators.

We can simplify this system by noting  that there are two functions $F(u_1,u_2,H,L_2)$, $G(u_1,u_2,H,L_2)$ such
that (\ref{partialxyHL}) is satisfied by 
\[
A=F,\qquad B=-\frac12 \partial_{2}F-\partial_{1}G,\qquad C=-\frac12\partial_{1} F+\partial_{2}G,
\]
Then the integrability condition for (\ref{partialxHL}), (\ref{partialyHL}) is (with the shorthand $\partial_{j}F=F_j$, $\partial_{j\ell}F=F_{j\ell}$, etc., for $F$ and $G$),
\begin{eqnarray} 2G_{1222}+\frac12 F_{2222}+2F_{22}(v_2-f_2H+L_2)
+3F_{2}(v'_2-f_2'H)+F(v''_2-f''_2H)&=&\nonumber\\  
 -2G_{1112} +\frac12 F_{1111}+2F_{11}(v_1-f_1H-L_2)
+3F_{1}(v'_1-f'_1H)+F(v''_1-f''_1H),&&\label{eqn1} \end{eqnarray} 
and equation (\ref{constanttermHL}) becomes
\begin{eqnarray}\label{eqn2} \frac12
F_{1112}+2F_{12}(v_1-f_1H)+F_{1}(v'_2-f'_2H)+\frac12
G_{1111}&+&\\
2G_{11}(v_1-f_1H-L_2)+G_{1}(v'_1-
f'_1H)&=&\nonumber\\
-\frac12F_{1222}-2F_{12}(v_2-f_2H)-F_{2}(v'_1-f'_1H)+\frac12
G_{2222}&+&\nonumber\\
2G_{22}(v_2-f_2H+L_2)+G_{2}(v'_2-f'_2H).&&\nonumber
\end{eqnarray}
We remark that   any solution of (\ref{eqn1}), (\ref{eqn2}) with $A,B,C$ not identically $0$ corresponds to a symmetry operator that does not commute with $L_2$, hence is algebraically  independent of the symmetries $H, L_2$.

 \section{ The potential $V=\alpha \frac{(x+iy)^{k-1}}{ (x-iy)^{k+1}}$}\label{sec3}
 
 We consider the flat space Schr\"odinger operator  
\be\label{generalk} H=\partial_{xx}+\partial_{yy}+V,\quad V=\alpha \frac{(x+iy)^{k-1}}{ (x-iy)^{k+1}},\ee
where $x,y$ are Cartesian coordinates.  We have shown that the corresponding classical systems are superintegrable for all rational $k$, \cite{KMPog10}.

In the special case $k=3$ we have explicitly established quantum superintegrability. Indeed, we 
obtained the  symmetry operators  
$$K_1=(\partial _x-i\partial _y)^3+ 
\frac{\alpha}{ (x-iy)^3}[-(iy+3x)\partial _x+(3iy+x)\partial _y],$$
$$K_2=(x\partial _y-y\partial _x)(\partial _x-i\partial _y)^3+ 
\frac{\alpha}{ (x-iy)^3}[i(2y^2-3ixy-3x^2)\partial ^2_x-(3iy+x)(iy+3x)\partial _x
\partial _y+$$
$$i(2x^2+3ixy-3y^2)\partial ^2_y-2i(3iy+x)\partial _x-2(iy+3x)\partial _y-8i]+i\alpha
^2 \frac{(x+iy)^3}{ (x-iy)^6},$$
$$K_3=(x\partial _y-y\partial _x)^2+2i\alpha y\frac {(-y^2+3x^2)}{ (x-iy)^3},$$
$$H=\partial ^2_x+\partial ^2_y+\alpha \frac {(x+iy)^6}{(x^2+y^2)^4},$$
with the commutation relations 
$$[K_1,K_2]=3iK^2_1,\ 
[K_1,K_3]=6iK_2-9K_1,$$
$$[K_2,K_3]=3i\{K_1,K_2\}+i(27+6\alpha )K_1+9K_2,$$
and the analogue of the constraint 
$$\frac{1}{ 2}\{K_1,K_1,K_3\}-3K^2_2-i\frac{9}{ 2}\{K_1,K_2\}+(\frac{63}{ 2} 
+3\alpha )K^2_1-3\alpha H^3=0.$$

All of these quantum systems separate in polar coordinates:
$$ u_1=R,\ u_2=\theta, \quad x=e^R\cos\theta,\ y=e^R\sin\theta.$$
The corresponding symmetry operator is 
$$-L_2\equiv K_2= \partial^2_\theta+\alpha e^{2ik\theta}.$$
Furthermore,
$$ H=e^{-2R}\left( \partial^2_R-L_2 \right),$$
and 
$$ f_1=e^{2R}, \ f_2=0,\ v_1=0,\ v_2=\alpha e^{2ik\theta}.$$
We assume $k=p/q$ for positive relatively prime integers $p,q$.
Based on the known expressions for the classical higher order constants of the motion, derived in \cite{}, we look for an operator constant of the motion
$\tilde L$, (\ref{generalLform}), where 
\be\label{FG}F=\sum_{a,b}{\cal A}_{a,b}(\alpha,H,L_2)e^{2(aR+ibk\theta)},\quad G=\sum_{a,b}{\cal B}_{a,b}(\alpha,H,L_2)e^{2(aR+ibk\theta)}.\ee
We require that there are only a finite number of nonzero terms in the sums and that the sums are of the form $a=a_0+m$, $b=b_0+n$ where $m,n$ run over a subset of the non-negative integers. (Thus ${\cal}C_{a_0,b_0}$ will be an analog of a lowest weight vector. Substituting all these expressions into equations (\ref{eqn1}), (\ref{eqn2}) and equating coefficients of terms $e^{2(aR+ibk\theta)}$, we obtain the matrix recursion 
\be\label{recursion1}  2(a^2-k^2b^2)
\left(\ba{cc} iakb& a^2+k^2b^2-L_2\\ -a^2-k^2b^2+L_2&4iakb\ea\right)\left(\ba{c} {\cal A}_{a,b}\\ {\cal B}_{a,b}\ea\right) +\ee
$$(2a-1)H\left( \ba{cc} -ikb& 1-a\\ a& 0\ea\right) \left( \ba{c} {\cal A}_{a-1,b} \\ {\cal B}_{a-1,b} \ea \right) +$$
$$
\alpha k (2b-1)\left(\ba{cc} ia& k(b-1)\\ -kb& 0\ea\right) \left(\ba{c} A_{a,b-1}\\ B_{a,b-1}\ea\right) =0,$$
or, solving for 
$$ {\cal C}_{a,b}=\left(\ba{c} {\cal A}_{a,b}\\ {\cal B}_{a,b}\ea\right),$$
\be\label{recursion2} {\cal C}_{a,b}+\frac{(2a-1)H}{J(a.b)}\left(\ba{cc} a(L_2+3k^2b^2-a^2)&4i(1-a)akb\\- ikb(L_2+k^2b^2)& -(a-1)(L_2+a^2+k^2b^2)\ea\right){\cal C}_{a-1,b}\ee
$$+\frac{\alpha k (2b-1)}{J(a,b)}\left(\ba{cc} -kb(L_2-k^2b^2+3a^2)&4i(b-1)ak^2b\\ ia(L_2+a^2)& k(b-1)(L_2+a^2+k^2b^2)\ea\right){\cal C}_{a,b-1}=0,$$
where
$$J(a,b)=2(a^2-k^2b^2)((a-kb)^2-L_2)((a+kb)^2-L_2).$$

Consider first the case where $p,q$ are both odd. We see from (\ref{recursion1}) that we can choose the 2-tuple ${\cal C}_{-p/2,q/2}$ arbitrarily, i.e., it is not a consequence of a recursion. Thus we set $a_0=-p/2$, $b_0=-q/2$, so that $a_0^2-k^2b_0^2=0$. Further we set ${\cal C}_{a,b}=0$ unless it can be computed explicitly from ${\cal C}_{a_0,b_0}$ by a sequence of recursions (\ref{recursion2}). 

Think of the elements ${\cal C}_{a,b}$ as laid out on an infinite grid, with rows labeled by $a$ and columns by $b$. The value of ${\cal C}_{a_0+m,b_0+n}$ for $m,\ne 0$ can be obtained via (\ref{recursion2}) as the sum of the contributions from all regular paths that lead from gridpoint $(a_0,b_0)$ to gridpoint ${a_0+m,b_0+n}$.  A regular path is a connected sequence of vertical moves upward: $({\tilde a}-1,{\tilde b})\to ({\tilde a},{\tilde b})$ and horizontal moves to the right: $({\tilde a},{\tilde b}-1)\to ({\tilde a},{\tilde b})$, in arbitrary order. All contributions from gridpoints below or to the left of $(a_0,b_0)$ are assumed zero. The contribution of a path to the value of ${\cal C}_{a_0+m,b_0+n}$ is the ordered product of the contributions of the vertical and horizontal segments that make up the regular path. 

We have assumed first that $p$ and $q$ are odd, so that $p/2,q/2$ are half-integers. Now notice that any regular path connecting $(a_0,b_0)=(-p/2,-q/2)$ to some gridpoint $(1/2, b)$ necessarily contains a vertical segment $(-1/2,{\tilde b})\to (1/2,{\tilde b})$ and contributes the factor 0, so that the contribution of the path to the sum is 0. Thus, necessarily, ${\cal C}_{1/2,b}=0$. Similarly, any regular path connecting $(a_0,b_0)=(-p/2,-q/2)$ to some gridpoint $(a, 1/2)$ necessarily contains an horizontal segment $({\tilde b},-1/2)\to ({\tilde b},1/2)$ and contributes the factor 0, so that the contribution of the path to the sum is 0. Thus,  ${\cal C}_{a,1/2}=0$. We conclude that the only possible nonzero 2-tuples are those in the grid $(-p/2+m,-q/2+n)$ where $0\le m<p/2$, $0\le n<q/2$, and these terms are uniquely determined by the choice of 
${\cal C}_{-p/2,-q/2}$. We get polynomial constants of the motion by taking the terms in ${\cal C}_{-p/2,-q/2}$ to be suitable finite products of the form
$$ \prod_{a,b} [J(a,b)],$$
to cancel the denominator terms in the the expressions for ${\cal C}_{a_0+m,b_0+n}$  which come from recursion (\ref{recursion2}). Thus we have constructed a 2-parameter family of finite order constants of the motion. It is a simple exercise to show that the commutators of these symmetries with $L_2$ are nonzero, so that the system is operator superintegrable. (Note that this last fact follows also from our original construction of the symmetries ${\tilde L}$. We must have $A\equiv B\equiv C=0$ unless ${\tilde L}$ is functionally independent of $H$ and $L_2$.)

If $k=-p/q$ with $p,q$ both odd, the same construction works with $(a_0,b_0)=(-p/2,-q/2)$.

The case $k=2^s p/q$ with $s\ge 1$ and $p, q$ relatively prime odd integers requires a modified analysis. Now we set $a_0=-2^{s-1}p$, $b_0=-q/2$, so that $a_0^2-k^2b_0^2=0$, $a_0$ is an integer and, as before, $b_0$ is half integer. Further we set ${\cal C}_{a,b}=0$ unless it can be computed explicitly from ${\cal C}_{a_0,b_0}$ by a sequence of recursions (\ref{recursion2}). As before, the value of ${\cal C}_{a_0+m,b_0+n}$ for $m,\ne 0$ can be obtained via (\ref{recursion2}) as the sum of the contributions from all regular paths that lead from gridpoint $(a_0,b_0)$ to gridpoint ${a_0+m,b_0+n}$. Now notice that  any vertical segment  connecting a gridpoint $(-1,{\tilde b})$ to gridpoint $(0,{\tilde b})$  maps ${\cal C}_{-1,{\tilde b}}$ to
$$ {\cal C}_{0,{\tilde b}}=\left(\ba{c} 0\\ {\tilde B}_{0,{\tilde b}}\ea\right),$$
i.e., to a 2-vector with upper component 0. Further, if this segment is followed by the horizontal segment connecting $(0,{\tilde b})$ to $(0,{\tilde b}+1)$ the upper component of the 2-vector will remain 0. Thus all regular paths that lead from gridpoint $(a_0,b_0)$ to any gridpoint $(0, b)$ on row $a=0$ will produce a 2-vector of the form
\be\label{special2vector}{\cal   C}_{0,{ b}}=\left(\ba{c} 0\\  B_{0,{ b}}\ea\right).\ee
Next, note that any vertical segment $(0,{\tilde b})\to (1,{\tilde b})$ will map a special 2-vector (\ref{special2vector}) to the zero vector. This means that ${\cal C}_{a,b}=0$ for all integers $a\ge 1$. Just as before, ${\cal C}_{a,b}=0$ for all half-integers $b\ge 1/2$.  Thus  the only possible nonzero 2-tuples are those in the grid $(-2^{s-1}p+m,-q/2+n)$ where $0\le m \le 2^{s-1}p$, $0\le n<q/2$, and these terms are uniquely determined by the choice of 
${\cal C}_{-p/2,-q/2}$. We get polynomial constants of the motion by taking the terms in ${\cal C}_{-p/2,-q/2}$ to be suitable finite products of the form
$$ \prod_{a,b} [J(a,b)],$$
to cancel the denominator terms in the the expressions for ${\cal C}_{a_0+m,b_0+n}$. Thus we have again constructed a 2-parameter family of finite order constants of the motion and the system is operator superintegrable. 

It is easy to extend these arguments to the cases $k=-2^{s}p/q$ and $k=\pm p/2^s q$ where $p,q$ are relatively prime odd integers. 
Thus the system (\ref{generalk}) is operator superintegrable for all rational $k$.

\section{The classical analog}
Here we first describe the classical analog of our infinite order symmetry operator construction and then apply it to the same example as in the previous section.  We construct constants of the motion of all orders for the  
Hamiltonian system 
\be \label{HS}{\cal H}=\sum_{j,k=1}^2 g^{jk}p_jp_k+V=E \ee
that admits a separation of variables. 
If $\{u_1,u_2\}$ defines  an orthogonal additive separable coordinate system for the Hamilton-Jacobi equation in 
some Riemannian space,  the corresponding Hamiltonian system
 has the form \cite{EIS34}
\begin{equation}
{\cal H}={\cal  L}_1= 
\frac{1}{f_1(u_1)+f_2(u_2)}\left(p^2_{u_1}+p^2_{u_2}+v_1(u_1)+v_2(u_2)
\right).\nonumber
\end{equation}
and, due to the separability, there is the second-order constant of the motion
\[
{\cal L}_2= \frac{f_2(u_2)}{f_1(u_1)+f_2(u_2)}\left(p^2_{u_1}+v_1(u_1)\right)-\frac{f_1(u_1)}{f_1(u_1)+f_2(u_2)}\left(p^2_{u_2}+v_2(u_2)\right),
\]
i.e., $
\{ {\cal L}_2, {\cal H}\}=0,$  where $\{\cdot,\cdot\}$ is the usual Poisson bracket, 
and we have  phase space identities 
\begin{equation} \label{fundident1}
f_1(u_1){\cal H} +{\cal L}_2=p^2_{u_1}+v_1(u_1),\qquad
f_2(u_2){\cal H}-{\cal L}_2=p^2_{u_2}+v_2(u_2).
\end{equation}

We  look for a  constant of the motion  ${ \cal \tilde L}({\cal H},{\cal L}_2,u_1,u_2)$, i.e., a function on the phase space  that satisfies 
\begin{equation}
\{{\cal H},{\cal \tilde  L}\}= 0.\label{newconditions2}
\end{equation}
We require that the constant of the motion take the standard form
\begin{equation}\label{standardLform1}
{\cal \tilde 
L}=\sum_{j,k}\left(A^{j,k}(u_1,u_2) p_{u_1}p_{u_2}+B^{j,k}(u_1,u_2)p_{u_1}
+C^{j,k}(u_1,u_2)p_{u_2}+ D^{j,k}(u_1,u_2)
\right){\cal H}^j{\cal L}_2^k.
\end{equation}
Note that if the formal symmetries  (\ref{standardLform1})
contained polynomial terms in
$p_{u_1}$ or $ p_{u_2}$  of orders $\ge 2$ we could use the  identities (\ref{fundident1}), recursively, and 
 rearrange terms to achieve the
unique standard form (\ref{standardLform1}).

We find that the symmetry condition (\ref{newconditions2}) is equivalent to the
system of equations
\begin{equation} \label{pxy}
\partial_{2}B^{j,k}+\partial_{1}C^{j,k}
=0,
\end{equation}
\be \label{px}
-2\partial_{2}A^{j,k}v_2+2\partial_{1}D^{j,k}-A^{j,k}v'_2+
2\partial_{2}A^{j-1,k}f_2+A^{j-1,k}f'_2-2\partial_{2}A^{j,k-1}=0,
\ee
\be \label{py}
-2\partial_{1}A^{j,k}v_1+2\partial_{2}D^{j,k}-A^{j,k}v'_1+
2\partial_{1}A^{j-1,k}f_1+A^{j-1,k}f'_1+2\partial_{1}A^{j,k-1}=0,
\ee
\begin{equation}\label{cterm}
-2\partial_{1}B^{j,k}v_1-2\partial_{2}C^{j,k}v_2-B^{j,k}v'_1-C^{j,k}v'_2 +2\partial_{1}B^{j-1,k}f_1
\end{equation}
\[
+2\partial_{2}C^{j-1,k}f_2+B^{j-1,k}f'_1
+C^{j-1,k}f'_2
+2\partial_{1}B^{j,k-1}-2\partial_{2}C^{j,k-1}=0.
\]

Note that condition (\ref{standardLform1}) makes sense, at least formally, for infinite order constants of the motion, and one can 
consider ${\cal H},{\cal L}_2$ as parameters in these equations. 

In this view we can write
\begin{eqnarray}\label{generalLform1}
{\cal \tilde
 L}({\cal H},{\cal L}_2,u_1,u_2)&=&A(u_1,u_2,{\cal H},{\cal L}_2)p_1p_2+B(u_1,u_2,{\cal H},{\cal L}_2)p_{1}\nonumber\\
&+&C(u_1,u_2,{\cal H},{\cal L}_2)p_{2}
+ D(u_1,u_2,{\cal H},{\cal L}_2),
\end{eqnarray}
and consider $\cal \tilde  L$ as an at most second-order  constant of the motion  that is analytic in the parameters ${\cal H},{\cal L}_2$. 
Then the above system of equations can be written in the more compact form
\begin{equation} \label{pxyHL}
B_{u_2}+C_{u_1}
=0,
\end{equation}
\begin{equation} \label{pxHL}
-2A_{u_2}v_2+2D_{u_1}-Av'_2+(2A_{u_2}f_2+Af'_2)H-2A_{u_2}L_2=0,
\end{equation}
\begin{equation} \label{pyHL}
-2A_{u_1}v_1+2D_{u_2}-Av'_1+(2A_{u_1}f_1+Af'_1)H+2A_{u_1}L_2=0,
\end{equation}
\begin{equation}\label{ctermHL}
-2B_{u_1}v_1-2C_{u_2}v_2-Bv'_1-Cv'_2
\end{equation}
\[
+(2B_{u_1}f_1+2C_{u_2}f_2+Bf'_1
+Cf'_2)H+(2B_{u_1}-2C_{u_2})L_2=0.
\]
We can view (\ref{pxyHL}) as an equation for $B,C$ and (\ref{pxHL}), (\ref{pyHL}) as the defining equations for $D_{u_1}, D_{u_2}$.

We can simplify this system, and easily compare it to the  operator system, by  noting  that there are two functions $F(u_1,u_2,{\cal H},{\cal L}_2)$, $G(u_1,u_2,{\cal H}, {\cal L}_2)$ such
that (\ref{pxyHL}) is satisfied by 
\[
A=F,\qquad B=-\partial_{1}G,\qquad C=\partial_{2}G,
\]
Then the integrability condition for (\ref{pxHL}), (\ref{pyHL}) is (with the shorthand $\partial_{j}F=F_j$, $\partial_{j\ell}F=F_{j\ell}$, etc., for $F$ and $G$),
\begin{eqnarray} 2F_{22}(v_2-f_2{\cal H}+{\cal L}_2)
+3F_{2}(v'_2-f_2'{\cal H})+F(v''_2-f''_2{\cal H})&=&\nonumber\\  
 2F_{11}(v_1-f_1{\cal H}-{\cal L}_2)
+3F_{1}(v'_1-f'_1{\cal H})+F(v''_1-f''_1{\cal H}),&&\label{eqn11} \end{eqnarray} 
and equation (\ref{ctermHL}) becomes
\begin{eqnarray}\label{eqn21} 
2G_{11}(v_1-f_1{\cal H}-{\cal L}_2)+G_{1}(v'_1-
f'_1{\cal H})&=&\\
2G_{22}(v_2-f_2{\cal H}+{\cal L}_2)+G_{2}(v'_2-f'_2{\cal H}).&&\nonumber
\end{eqnarray}

 Now we use this classical construction to study the flat space Hamiltonian system 
\be\label{gk} {\cal H}=p_x^2+p_y^2+V,\quad V=\alpha \frac{(x+iy)^{k-1}}{ (x-iy)^{k+1}},\ee
where $x,y$ are Cartesian coordinates.  We have already shown that this system is superintegrable for all rational $k$, \cite{KMPog10}.

All of these classical  systems separate in polar coordinates:
$$ u_1=R,\ u_2=\theta, \quad x=e^R\cos\theta,\ y=e^R\sin\theta,$$
with corresponding constants of the motion
$$-{\cal L}_2= p^2_\theta+\alpha e^{2ik\theta}.$$
Furthermore,
$${\cal  H}=e^{-2R}\left( p^2_R-{\cal L}_2 \right),$$
and 
$$ f_1=e^{2R}, \ f_2=0,\ v_1=0,\ v_2=\alpha e^{2ik\theta}.$$
We assume $k=p/q$ for  relatively prime integers $p,q$.
Based on the known expressions for the classical higher order constants of the motion, derived in \cite{KMPog10}, we look for a standard form  constant of the motion
$\cal \tilde L$, (\ref{generalLform1}), where 
\be\label{FG1}F=\sum_{a,b}{\cal A}_{a,b}(\alpha,{\cal H},{\cal L}_2)e^{2(aR+ibk\theta)},\quad G=\sum_{a,b}{\cal B}_{a,b}(\alpha,{\cal H},{\cal L}_2)e^{2(aR+ibk\theta)}.\ee
We require that there are only a finite number of nonzero terms in the sums and that the sums are of the form $a=a_0+m$, $b=b_0+n$ where $m,n$ run over a subset of the non-negative integers. Substituting all these expressions into equations (\ref{eqn11}), (\ref{eqn21}) and equating coefficients of terms $e^{2(aR+ibk\theta)}$, we obtain the matrix recursion 
\be\label{recursion3}  2{\cal L}_2(a^2-k^2b^2)\left(\ba{c} {\cal A}_{a,b}\\ {\cal B}_{a,b}\ea\right) +
(2a-1)H\left(\ba{cc} a&0\\ 0& a-1\ea\right)\left(\ba{c} {\cal A}_{a-1,b}\\ {\cal B}_{a-1,b}\ea\right)\ee
$$-\alpha k^2 (2b-1)\left(\ba{cc} b&0\\ 0& b-1\ea\right)\left(\ba{c} {\cal A}_{a,b-1}\\ {\cal B}_{a,b-1}\ea\right).$$
Although this system of equations is much simpler than the corresponding operator equations (\ref{recursion1}), (\ref{recursion2}), it shares essential features with them so that the details of the proof of superintegrability are essentially unchanged. As before we set 
$$ {\cal C}_{a,b}=\left(\ba{c} {\cal A}_{a,b}\\ {\cal B}_{a,b}\ea\right).$$

Consider first the case where $p,q$ are both odd and positive. We see from (\ref{recursion3}) that we can choose the 2-tuple ${\cal C}_{-p/2,q/2}$ arbitrarily.  Thus we set $a_0=-p/2$, $b_0=-q/2$, so that $a_0^2-k^2b_0^2=0$ and we set ${\cal C}_{a,b}=0$ unless it can be computed explicitly from ${\cal C}_{a_0,b_0}$ by a sequence of recursions (\ref{recursion3}). 

 The value of ${\cal C}_{a_0+m,b_0+n}$ for $m,\ne 0$ can be obtained via (\ref{recursion3}) as the sum of the contributions from all regular paths that lead from gridpoint $(a_0,b_0)$ to gridpoint ${a_0+m,b_0+n}$.  Since  $p/2,q/2$ are half-integers,  any regular path connecting $(a_0,b_0)=(-p/2,-q/2)$ to some gridpoint $(1/2, b)$ necessarily contains a vertical segment $(-1/2,{\tilde b})\to (1/2,{\tilde b})$ and contributes the factor 0, so that the contribution of the path to the sum is 0. Thus, necessarily, ${\cal C}_{1/2,b}=0$. Similarly, any regular path connecting $(a_0,b_0)=(-p/2,-q/2)$ to some gridpoint $(a, 1/2)$ necessarily contains an horizontal segment $({\tilde b},-1/2)\to ({\tilde b},1/2)$ and contributes the factor 0, so that the contribution of the path to the sum is 0. Thus,  ${\cal C}_{a,1/2}=0$. We conclude that the only possible nonzero 2-tuples are those in the grid $(-p/2+m,-q/2+n)$ where $0\le m<p/2$, $0\le n<q/2$, and these terms are uniquely determined by the choice of 
${\cal C}_{-p/2,-q/2}$. We get polynomial constants of the motion by taking the terms in ${\cal C}_{-p/2,-q/2}$ to be suitable powers of ${\cal L}_2$ 
to cancel the denominator terms in the the expressions for ${\cal C}_{a_0+m,b_0+n}$  which come from recursion (\ref{recursion3}). Thus we have constructed a 2-parameter family of finite order constants of the motion. It is easy to show that the Poisson brackets  of these symmetries with ${\cal L}_2$ are nonzero, so that the system is classically superintegrable.  There is a special simplification here in that the recursion (\ref{recursion3}) decouples into separate equations for ${\cal A}_{a,b}$ and for ${\cal B}_{a,b}$. 

If $k=-p/q$ with $p,q$ both odd, the same construction works with $(a_0,b_0)=(-p/2,-q/2)$.

For case $k=2^s p/q$ with $s\ge 1$ and $p, q$ relatively prime odd positive integers, we set $a_0=-2^{s-1}p$, $b_0=-q/2$, so that $a_0^2-k^2b_0^2=0$, $a_0$ is an integer and, as before, $b_0$ is half integer. Again we set ${\cal C}_{a,b}=0$ unless it can be computed explicitly from ${\cal C}_{a_0,b_0}$ by a sequence of recursions (\ref{recursion3}). As before, the value of ${\cal C}_{a_0+m,b_0+n}$ for $m,\ne 0$ can be obtained via (\ref{recursion3}) as the sum of the contributions from all regular paths that lead from gridpoint $(a_0,b_0)$ to gridpoint ${a_0+m,b_0+n}$. However, any vertical segment  connecting a gridpoint $(-1,{\tilde b})$ to gridpoint $(0,{\tilde b})$  maps ${\cal C}_{-1,{\tilde b}}$ to
$$ {\cal C}_{0,{\tilde b}}=\left(\ba{c} 0\\ {\tilde B}_{0,{\tilde b}}\ea\right),$$
i.e., to a 2-vector with upper component 0. If this segment is followed by the horizontal segment connecting $(0,{\tilde b})$ to $(0,{\tilde b}+1)$ the upper component of the 2-vector will remain 0, so  all regular paths that lead from gridpoint $(a_0,b_0)$ to any gridpoint $(0, b)$ on row $a=0$ will produce a 2-vector of the form
\be\label{special2vector1}{\cal   C}_{0,{ b}}=\left(\ba{c} 0\\  B_{0,{ b}}\ea\right).\ee
Note that any vertical segment $(0,{\tilde b})\to (1,{\tilde b})$ will map a special 2-vector (\ref{special2vector1}) to the zero vector. Thus ${\cal C}_{a,b}=0$ for all integers $a\ge 1$, and as before, ${\cal C}_{a,b}=0$ for all half-integers $b\ge 1/2$.  Thus  the only possible nonzero 2-tuples are those in the grid $(-2^{s-1}p+m,-q/2+n)$ where $0\le m \le 2^{s-1}p$, $0\le n<q/2$, and these terms are uniquely determined by the choice of 
${\cal C}_{-p/2,-q/2}$. We get polynomial constants of the motion by taking the terms in ${\cal C}_{-p/2,-q/2}$  to be suitable powers of ${\cal L}_2$ 
to cancel the denominator terms in the the expressions for ${\cal C}_{a_0+m,b_0+n}$. We have again constructed a 2-parameter family of finite order constants of the motion and the system is classically superintegrable. There is a significant simplification here, due to the decoupling of (\ref{recursion3})  into separate equations for ${\cal A}_{a,b}$ and for ${\cal B}_{a,b}$. If we choose ${\cal B}_{a_0,b_0}=0$ then all vectors ${\cal C}_{0,b}=0$, so that row 0 can be removed from the grid.

Again it is easy to extend these arguments to the cases $k=-2^{s}p/q$ and $k=\pm p/2^s q$ where $p,q$ are relatively prime odd integers. 
Thus we have a new proof that the system (\ref{gk}) is classically superintegrable for all rational $k$, and have clarified the relation between the classical and operator symmetries for this system.

{\bf Example}: For system (\ref{gk}) with $k=3$ the recurrence relations can be solved easily to give 
$$F={\cal L}_2 e^{-3R-3i\theta}+\frac14 {\cal H}e^{-R-3i\theta},\  G={\cal L}_2  e^{-3R-3i\theta}+\frac34 {\cal H} e^{-R-3i\theta},$$
so
$$  A=F,\ B=3{\cal L}_2 e^{-3R-3i\theta}+\frac34 {\cal H}e^{-R-3i\theta},\ C=-3i{\cal L}_2 e^{-3R-3i\theta}-\frac94 {\cal H} e^{-R-3i\theta},\ D=\frac{i}{4}{\cal L}_2 (4{\cal L}_2+3e^{2R} {\cal H})e^{-3R-3i\theta}.$$  
This gives us the canonical forms for the  3rd and 4th order constants of the motion ${\cal K}_1,\ {\cal K}_2$ as given in \cite{KMPog10}. Indeed,
$A p_R p_\theta+D=\frac34 {\cal K}_1$,  $B p_R+C p_\theta=\frac14 {\cal K}_2$.

\section{The TTW system}
Now we apply our constructions to the TTW system, \cite{TTW, TTW2}.  Here,
\be\label{TTW} u_1=R,\ u_2=\theta,\ f_1=e^{2R},\ f_2=0,\ v_1=\alpha e^{4R},\ee
$$ v_2=\frac{\beta}{\cos^2(k\theta)}+\frac{\gamma}{\sin^2(k\theta)}=\frac{2(\gamma+\beta)}{\sin^2(2k\theta)}+\frac{2(\gamma-\beta)\cos(2k\theta)}{\sin^2(2k\theta)}.$$
Based on the results of \cite{KMPog10} for the classical case, we postulate expansions of  $F,G$ in finite series
\be\label{TTW1} F=\sum_{a,b,c} A_{a,b,c}E_{a,b,c}(R,\theta),\quad G=\sum_{a,b,c} B_{a,b,c}E_{a,b,c}(R,\theta),\ee
$$ E_{a,b,0}=e^{2aR}\sin^b(2k\theta),\quad E_{a,b,1}=e^{2aR}\sin^b(2k\theta)\cos(2k\theta).$$
The sum is taken over terms of the form $a=a_0+m$, $b=b_0+n$, and $c=0,1$, where $m,n$ are integers.  The point $(a_0,b_0)$ could in principle be any point in $\mathbb R^2$, however,
for reasons discussed below, we will take $a_0$ to be a positive integer and $b_0$ to be a negative integer.

Taking coefficients with respect to the basis (\ref{TTW1}) in each of equation (\ref{eqn1}) and (\ref{eqn2})
gives recurrence relations for these coefficients.  For example, the coefficient of
$e^{2aR}\sin^{b-2}(2k\theta)\cos(2k\theta)$ in equation (\ref{eqn1}) gives the equation
\[
8b{k}^{2} \left( b-1 \right)  \left( L_2 - 2 ({k}^{2}(b^2+1)+\gamma+\beta) \right) A_{{a,b,1}}
\ + \ 32 a{k}^{3}b \left( b^2-1 \right) B_{{a,b+1,0}}
\]
\[
{} \ + \ 8 {k}^{2} \left( b^2-1 \right) 
      \left( {b}^{2}{k}^{2}+2 b{k}^{2}+2 \beta+2 \gamma \right) A_{{a,b+2,1}}
\ + \ 16 {k}^{2} \left( b^2-1 \right)  \left( \gamma-\beta \right) A_{{a,b+2,0}}
\]
\[
{} \ - \ 8 {k}^{2} \left( 2 b-1 \right)  \left( b-1 \right)  \left( \gamma-\beta \right) A_{{a,b,0}}
 \ + \ 8  \left( a^2-k^2(b-1)^2 \right)  \left( L_2-{a}^{2}- k^2(b-1)^2 \right) A_{{a,b-2,1}}
\]
\[
{} \ + \ 4 H a \left( 2 a-1 \right) A_{{a-1,b-2,1}}-8 
a\alpha  \left( a-1 \right) A_{{a-2,b-2,1}}
 \ + \ 32 ak \left( b-1 \right)  \left( a^2-k^2(b-1)^2 \right) B_{{a,b-1,0}} \ = \ 0.
\]
The shifts in the indices of $A$ and $B$ are integers and so we can view this as an equation on a two-dimensional
lattice with integer spacings.  While the shifts in the indices are of integer size, we haven't required that the 
indices themselves be integers, although they may be integers in particular examples.

Taking the coefficient of $e^{2aR}\sin^b(2k\theta)$ in equation (\ref{eqn1}) 
and the coefficients of $e^{2aR}\sin^{b-1}(2k\theta)\cos(2k\theta)$ and
$e^{2aR}\sin^{b-1}(2k\theta)$ in equation (\ref{eqn2}) give a further three recurrence relations.  At a general
point in the lattice there are 4 coefficients, and these 4 equations will be shown to be independent.
The equations are linear and homogeneous and so there must be some points where the independence of
the equations breaks down and allows at least one coefficient to be arbitrarily chosen.

The different powers of $\sin(2k\theta)$ used in obtaining these equations have
been chosen as a matter of convenience
after many experiments conducted using the computer algebra package Maple.

All four recurrence relations are of a similar complexity, but rather than write them
out separately, we will combine them into a matrix recurrence relation by
defining
\[
 {C}_{a,b} = \left( \begin{array}{c} A_{a,b,0} \\ B_{a,b-1,0} \\ A_{a,b-2,1} \\ B_{a,b-1,1} \end{array} \right).
\]
We can now write the 4 recurrence relations in matrix form as
\begin{eqnarray}
 \mathbf{0} &=& {M}_{a,b}{C}_{a,b}  + 
 {M}_{a,b-2}{C}_{a,b-2} + 
 {M}_{a,b-4}C_{a,b-4} + 
 {M}_{a,b-6}C_{a,b-6} \nonumber \\
 & & {} \quad + 
 {M}_{a-1,b}{C}_{a-1,b} + 
 {M}_{a-1,b+2}{C}_{a-1,b+2} + 
 {M}_{a-2,b}C_{a-2,b} + 
 {M}_{a-2,b+2}{C}_{a-2,b+2} \label{matrixrec},
\end{eqnarray}
where each ${M}_{i,j}$ is a $4\times4$ matrix given below.  It is useful to visualize the
the set of points in the lattice which enter into this recurrence for a given choice of $(a,b)$.
These are represented in figure \ref{template} in which the upper left corner is the point $(a,b)$.
From this it is clear that when ${M}_{a,b}$
is nonsingular, the value of ${C}_{a,b}$ can be uniquely determined from the 8 points to
its right and below.  In that case,
\begin{eqnarray}
 {C}_{a,b}  &=& -{M}_{a,b}^{-1}\Bigl( 
 {M}_{a,b-2}{C}_{a,b-2} + 
 {M}_{a,b-4}{C}_{a,b-4} + 
 {M}_{a,b-6}{C}_{a,b-6} \nonumber \\
 & & {} \quad + 
 {M}_{a-1,b}{C}_{a-1,b} + 
 {M}_{a-1,b+2}{C}_{a-1,b+2} + 
 {M}_{a-2,b}{C}_{a-2,b} + 
 {M}_{a-2,b+2}{C}_{a-2,b+2} \Bigr). \label{Cabrec}
\end{eqnarray}

This allows us to construct an iterative procedure that calculates the values of
${C}_{a,b}$ at points in the lattice using only other points where
the values of ${C}_{i,j}$ are already known.  Since the point $(a,b)$
corresponds to the top left corner of the collection of points in Figure \ref{template},
this process will calculate the coefficients in a sequence that moves from right
to left and bottom to top.  Note the matrices
corresponding to the right hand `corners' of the set of points in Figure \ref{template}
(${M}_{a,b+6}$ and ${M}_{a-2,b+2}$) are singular and so could not be used 
in the same way, while the matrix corresponding
to the bottom left corner is generically not singular, however it has properties that
we will use for another purpose.

\begin{figure}
\[
 \begin{array}{ccccccccccc}
 \cdot & \cdot   & \cdot & \cdot   & \cdot & \cdot   & \cdot & \cdot   & \cdot & \cdot   & \cdot \\
 \cdot & \bullet & \cdot & \bullet & \cdot & \bullet & \cdot & \bullet & \cdot & \bullet & \cdot \\
 \cdot & \bullet & \cdot & \bullet & \cdot & \cdot   & \cdot & \cdot   & \cdot & \cdot   & \cdot \\
 \cdot & \bullet & \cdot & \bullet & \cdot & \cdot   & \cdot & \cdot   & \cdot & \cdot   & \cdot \\
 \cdot & \cdot   & \cdot & \cdot   & \cdot & \cdot   & \cdot & \cdot   & \cdot & \cdot   & \cdot \\
 \end{array}
\] 
\caption{Points contributing to the recurrence relation are marked with large dot ($\bullet$).  The large dot
in the upper
left corner corresponds to the position $(a,b)$.}
\label{template}
\end{figure}

%
%

\[
{M}_{a,b} =
 \left( \begin {array}{cc}
 8 {k}^{2} \left( 2 b-1 \right) \left( b-1 \right)  \left( \beta-\gamma \right)
 & 32 ak \left( b-1 \right)  \left( a^2-k^2(b-1)^2 \right) \\
 8  \left( a^2-k^2b^2\right)  \left( L_2 -{a}^{2} -{b}^{2}{k}^{2} \right)
 & 0 \\
 8 abk \left( a^2-k^2b^2 \right) 
 & 0 \\
 8 ak \left( 2 b-1 \right)  \left( \beta-\gamma \right)
 & -8  \left( a^2-k^2(b-1)^2 \right)  \left(  L_2-{a}^{2} - k^2(b-1)^2 \right)
 \end {array} \right.
\]
\[
 \left. \begin {array}{cc}
 8  \left( a^2-k^2(b-1)^2 \right)  \left(  L_2-{a}^{2} -  k^2(b-1)^2 \right)
 & 0 \\
 0
 & -32 abk \left( a^2-k^2b^2 \right) \\
 0 
 & -8  \left( a^2-k^2b^2 \right)  \left(  L_2-{a}^{2} -{b}^{2}{k}^{2} \right) \\
 -8 ak \left( b-1 \right)  \left( a^2-k^2(b-1)^2 \right)
 & 8 b{k}^{2} \left( 2 b-1 \right) \left( \beta-\gamma \right)
 \end {array} \right) 
\]

%
%

\[
{M}_{a,b+2} = 
 \left( \begin {array}{cc}
 16 {k}^{2} \left( b^2-1 \right)  \left( \gamma-\beta \right)
 & 32 a{k}^{3}b \left( b^2-1 \right) \\
%
%
 8 \left(b+1\right)k^2\left((b+2)(L_2 -2k^2(b^2+2b+2)) -2b(\beta+\gamma ) \right)
 & 0 \\
 8 ak \left( b+1 \right)  \left(  k^2(b+2)  +2 (\beta+ \gamma) \right)
 & 8k^2(b+1)(2b+1)(\gamma-\beta) \\
 16 ak \left( b+1 \right)  \left( \gamma-\beta \right)
 & \hspace{-10mm} 8 b{k}^{2} \left( b+1 \right)  \left( 2(k^2(b^2+1) +\beta+\gamma) - L_2  \right)
 \end {array}
 \right.
\]

\[
 \left. \begin {array}{cc}
 8 b{k}^{2} \left( b-1 \right)  \left( L_2 -2(k^2(b^2+1)+\beta+\gamma)  \right)
 & 0 \\
 8 b{k}^{2} \left( 2b+1 \right) \left( \gamma-\beta \right)
 & 32 ak \left( b+1 \right)  \left( {a}^{2} -2k^2(b+2)^2 \right) \\
 8 ak \left( 2b+1 \right)  \left( \beta-\gamma \right)
 & 8  \left( b+1 \right) {k}^{2} \left(2(b+2)(\beta+\gamma) + 2bk^2(b^2+2b+2) - bL_2  \right) \\
 8 abk \left( {a}^{2} - 2(k^2(b^2+1)+\beta+\gamma)  \right)
 & 8 {k}^{2} \left( b+1 \right) 
 \left( 4 b+3 \right)  \left( \gamma-\beta \right) \end {array}
 \right) 
\]

%
%

\[
{M}_{a,b+4} = 
 \left( \begin {array}{cc}
 0
 &
 0 \\
 8 {k}^{2} \left( b+3 \right)  \left( b+1 \right)  \left( k^2(b+4)(b+2) +2( \beta+\gamma) \right)
 & 0 \\
 0
 & 16 {k}^{2} \left( b+3 \right)  \left( b+1 \right)  \left(\beta-\gamma\right) \\
 0
 & -8 {k}^{2} \left( b+3 \right)  \left( b+1 \right)  \left( bk^2(b+2) + 2(\beta+\gamma) \right)
 \end {array} \right.
\]
\[
 \left. \begin {array}{cc}
 8 {k}^{2} \left( b-1 \right) \left( b+1 \right)  \left(  bk^2(b+2) + 2(\beta+\gamma) \right)
 & 0 \\
 8 {k}^{2} \left( 4 b+5 \right)  \left( b+1 \right)  \left( \beta-\gamma \right)
 & 32 a{k}^{3} \left( b+3 \right)  \left( b+2 \right)  \left( b+1 \right) \\
 16 ak \left( b+1 \right)  \left( \gamma-\beta \right) 
 & -8 {k}^{2} \left( b+3 \right)  \left( b+1 \right)  \left(  bk^2(b+2) + 2(\beta+\gamma) \right) \\
 8 ak \left( b+1 \right)  \left(  bk^2(b+2) + 2(\beta+\gamma) \right)
 & 16 {k}^{2} \left( b+3 \right)  \left( b+1 \right)  \left( \beta-\gamma \right)
 \end {array} \right) 
\]

\[
{M}_{a,b+6} = 
 \left( \begin {array}{cccc} 0&0&0&0\\ \noalign{\medskip}0&0&16 {k}^{
2} \left( b+3 \right)  \left( b+1 \right)  \left( \gamma-\beta
 \right) &0\\ \noalign{\medskip}0&0&0&0\\ \noalign{\medskip}0&0&0&0
\end {array} \right) 
\]

\[
{M}_{a-1,b} = 
 \left( \begin {array}{cccc} 0&0&4  H  a \left( 2 a-1 \right) 
&0\\ \noalign{\medskip}4  H  a \left( 2 a-1 \right) &0&0&0
\\ \noalign{\medskip}-4  H  bk \left( 2 a-1 \right) &0&0&-4 H  \left( 2 a-1 \right)  \left( a-1 \right) 
\\ \noalign{\medskip}0&-4  H   \left( 2 a-1 \right)  \left( a-
1 \right) &4  H  k \left( b-1 \right)  \left( 2 a-1 \right) &0
\end {array} \right) 
\]

\[
{M}_{a-1,b+2} =  \left( \begin {array}{cccc}
 0&0&0&0\\0&0&0&0\\0&0&0&0\\ 0&0&-4  H  bk \left( 2 a-1 \right) &0
\end {array} \right) 
\]

\[
{M}_{a-2,b} =  \left( \begin {array}{cccc}
 0&0&-8 a\alpha  \left( a-1 \right) &0 \\
 -8 a\alpha  \left( a-1 \right) &0&0&0 \\
 8 \alpha kb \left( a-1 \right) &0&0&8 \alpha  \left( a-1 \right)  \left( a-2 \right) \\
 0&8 \alpha  \left( a-1 \right)  \left( a-2 \right) &-8 \alpha k \left( b-1 \right)  \left( a-1 \right) &0
\end {array} \right) 
\]

\[
{M}_{a-2,b+2} =  \left( \begin {array}{cccc} 0&0&0&0\\ \noalign{\medskip}0&0&0&0
\\ \noalign{\medskip}0&0&0&0\\ \noalign{\medskip}0&0&8 \alpha kb
 \left( a-1 \right) &0\end {array} \right) 
\]

We are interested in finding a solution to the recurrence relation that gives $F$ and $G$ as finite
sums and hence we seek solutions that are confined to a finite rectangle in the lattice.  Since the
equations for the $A_{i,j,k}$ and $B_{i,j,k}$ are linear and homogeneous, they always admit the
trivial solution and so we need to demonstrate that a nonzero solution can be found.  Our
approach is as follows.

For a finite solution, there must be a lowest nonzero row and in that row, a rightmost
nonzero element.  Label this rightmost point in the bottom row as $(a_0,b_0)$.
Since all elements to the right and below this point are zero, ${M}_{a_0,b_0}$
must be singular, otherwise we could use (\ref{Cabrec}) to show that ${C}_{a_0,b_0}$
must vanish, contradicting its definition.

Since
\[
\mbox{det}({M}_{a,b}) =  -4096 \left( a^2-k^2b^2 \right) ^{2} \left( a^2-k^2(b-1)^2\right) ^{2}
 \left( L_2-(a+k(b-1))^2 \right)
 \left( L_2-(a-k(b-1))^2 \right) \times
\]
\begin{equation} 
\label{detMab} {} \qquad\qquad\qquad  \times
 \left( L_2-(a+kb)^2 \right)
 \left( L_2-(a-kb)^2 \right)
 \end{equation}
we must choose our starting point so that either $a^2=k^2b^2$ or $a^2=k^2(b-1)^2$, that is, if
$k=p/q$ with $p$ and $q$ a pair of relatively prime positive integers, we can choose $(a_0,b_0)$ to
be one of $(\eta p, \eta q)$, $(-\eta p, \eta q)$, $(\eta p,\eta q+1)$ or $(\eta p,-\eta q+1)$ for
any real number $\eta$.
At all of these points, the rank of ${M}_{a_0,b_0}$ is 2 and hence at these points we can choose 
$4-\mbox{Rank}({M}_{a,b})=2$ components of ${C}_{a,b}$ to be arbitrary parameters.

In order to have a finite solution, 
we must eventually reach a point in the lattice beyond which all entries vanish or can be chosen to vanish.
Examining the matrices defining the recurrence, we see that it may be possible to achieve this a left
hand boundary due to the
many terms with factors such as $b-1$, $b+1$, $b+3$ and on an upper boundary because of factors of $a$ and $a-1$.
For this reason, we will now take $a_0$ to be $-p$ and $b_0$ to be $q$ or $q+1$ 
and examine how the cut offs on the left and top occur.

As the recurrence relations (\ref{matrixrec}) and (\ref{Cabrec}) only involve shifts of multiples of two units
in second index, it is easy to verify that all entries in columns that are an odd number of steps away from column 
$b_0$ must vanish or can be chosen to vanish.  Furthermore, we have two candidates for $b_0$, $q$ and $q+1$.  We
will choose $b_0$ to be which ever of these is odd.
We can then assume that even numbered columns have only vanishing entries and can traverse the lattice in 
steps of two to the left
starting from column $b_0$.

We now work our way across row $a_0$ starting from column $b_0$ taking steps of two units to the left.
At the first point, $(a,b)=(a_0,b_0)$,  ${M}_{a,b}$ has rank 2 and so the components of
${C}_{a,b}$ depend linearly on two arbitrary parameters.  At other points in the bottom row,
${M}_{a,b}$ is nonsingular (unless we reach $b=-b_0$) and so we can solve for $C_{a_0,b}$.
At the points when $b=1,-1,-3$, this takes a special form.  Note that in the bottom row, all lower points have vanishing $C_{i,j}$ and so we need only
consider contributions from the points  $(a,b+2)$, $(a,b+4)$ and $a,b+6)$. We will initially assume that $q\neq 1,2,3,4,5,6$ so that ${M}_{-p,1}$, ${M}_{-p,-1}$, ${M}_{-p,-3}$ and ${M}_{-p,-5}$ are all nonsingular.  This
is not essential, however the argument is simpler in this case.

First consider $b=1$.   The form of the matrices giving contributions from points to the right are the first two matrices in Table \ref{leftboundary}.  It is clear from these that the third component of ${C}_{a_0,1}$ must vanish.  Next consider $b=-1$.
The only nonzero matrix elements occur in column 3 of ${M}_{a_0,-1}^{-1}{M}_{a_0,1}$ and so
\[
 {C}_{a_0,-1} = {M}_{a_0,-1}^{-1}{M}_{a_0,1}{C}_{a_0,1}
 + {M}_{a_0,-1}^{-1}{M}_{a_0,3}{C}_{a_0,3}
 + {M}_{a_0,-1}^{-1}{M}_{a_0,5}{C}_{a_0,5} = \mathbf 0.
\]
A similar calculation shows that ${C}_{a_0,-3}={C}_{a_0,-5} = \mathbf 0$ and hence 
${C}_{a_0,j} = \mathbf 0$ for all $j\leq-1$.

Next we repeat the process for the row above, that is, row $a_0+1$ starting from the right hand end, and then again for
row $a_0+2$ and so.  The argument showing that all ${C}_{i,j}$ vanish for $j\leq-1$ is essentially the same
as for row $a_0$ except there are a few extra terms to consider since the elements in the row below are no longer all zero.

To see how the cut off occurs at the top, start at the right hand end of the $0$ row, that is in position $(0,b_0)$.
All elements to the right are zero and so the only contributions to  ${C}_{0,b_0}$ come from below, that is
from $(-1,b_0)$ and $(-2,b_0)$.  It is clear from the corresponding matrices in Table \ref{topboundary} that
the first and third components of ${C}_{0,b_0}$ are zero.  Stepping across the row in step of two to the left,
it is easy to check that this is maintained for all elements of this row.

Next consider row $1$.  From the form of the matrices given for $a=1$ in Table \ref{topboundary} it 
is clear that the only the first and third components of ${C}_{0,j}$ can contributed to any ${C}_{1,j'}$.
However, since these components have already been shown to vanish we conclude that ${C}_{1,j}=\mathbf 0$ for
all $j$.

The last step is to consider row $2$.  As for row $1$, when $a=2$ the form of the matrices given for $a=2$ in Table \ref{topboundary} clearly shows that only the first and third components of ${C}_{0,j}$ can contributed to any ${C}_{2,j'}$.
As these components have already been shown to vanish we conclude that ${C}_{2,j}=\mathbf 0$ for
all $j$.

Since we have two completely zero rows, it is now clear that ${C}_{i,j}=\mathbf 0$ for all $i\geq1$.

The above argument needs modification to see that the left hand cut off can be achieved 
when treating the bottom row for $q=1,2,3,4,5$ or $6$ as
${M}_{a,b}$ will be singular in one of the columns $b=-1$, $-3$ or $-5$.  However, it is
a simple matter to use the original matrix recurrence relations (\ref{matrixrec}) to check that
the same conclusions can be reached in each of these cases, that is, the third component
of ${C}_{a_0,1}$ vanishes and each entry of ${C}_{a_0,j}$ for $j\leq 1$ 
is either required to vanish or can be chosen to vanish.

\begin{table}
\begin{tabular}{ccccc}
$b=1$:
  & $\left(\begin{array}{cccc} *&*&*&* \\ *&*&*&* \\ 0&0&0&0 \\ *&*&*&* \end{array}\right)$ 
  & $\left(\begin{array}{cccc} 0&0&*&0 \\ 0&0&*&0 \\ 0&0&0&0 \\ 0&0&*&0 \end{array}\right)$ 
  & $\left(\begin{array}{cccc} *&0&0&* \\ *&*&0&* \\ 0&0&*&0 \\ *&0&0&* \end{array}\right)$ 
  & $\left(\begin{array}{cccc} 0&0&0&0 \\ 0&0&*&0 \\ 0&0&0&0 \\ 0&0&0&0 \end{array}\right)$ \\[27pt]
  & ${M}_{a,b}^{-1}{M}_{a,b+2}$ & ${M}_{a,b}^{-1}{M}_{a,b+6}$ 
     & ${M}_{a,b}^{-1}{M}_{a-1,b}$  
     & ${M}_{a,b}^{-1}{M}_{a-1,b+2}$ \\
  & ${M}_{a,b}^{-1}{M}_{a,b+4}$ & & ${M}_{a,b}^{-1}{M}_{a-2,b}$
     & ${M}_{a,b}^{-1}{M}_{a-2,b+2}$\\[20pt]
$b=-1$:
  & $\left(\begin{array}{cccc} 0&0&*&0 \\ 0&0&*&0 \\ 0&0&*&0 \\ 0&0&*&0 \end{array}\right)$ 
  & $\left(\begin{array}{cccc} 0&0&0&0 \\ 0&0&0&0 \\ 0&0&0&0 \\ 0&0&0&0 \end{array}\right)$ 
  & $\left(\begin{array}{cccc} 0&0&0&0 \\ 0&0&*&0 \\ 0&0&*&0 \\ 0&0&0&0 \end{array}\right)$ 
  &  \\[27pt]
  & ${M}_{a,b}^{-1}{M}_{a,b+2}$ & ${M}_{a,b}^{-1}{M}_{a,b+4}$
   & ${M}_{a,b}^{-1}{M}_{a-1,b+2}$ & \\
  & & ${M}_{a,b}^{-1}{M}_{a,b+6}$ & ${M}_{a,b}^{-1}{M}_{a-2,b+2}$\\[20pt]
$b=-3$:
  & $\left(\begin{array}{cccc} 0&0&*&0 \\ 0&0&*&0 \\ 0&0&*&0 \\ 0&0&*&0 \end{array}\right)$ 
  & $\left(\begin{array}{cccc} 0&0&0&0 \\ 0&0&0&0 \\ 0&0&0&0 \\ 0&0&0&0 \end{array}\right)$ 
  & 
  &  \\[27pt]
  & ${M}_{a,b}^{-1}{M}_{a,b+4}$ & ${M}_{a,b}^{-1}{M}_{a,b+6}$ & & \\[20pt]
$b=-5$:
  & $\left(\begin{array}{cccc} 0&0&*&0 \\ 0&0&*&0 \\ 0&0&*&0 \\ 0&0&*&0 \end{array}\right)$ 
  & 
  & 
  &  \\[27pt]
  & ${M}_{a,b}^{-1}{M}_{a,b+6}$ & & & \\[20pt]
\end{tabular} 
\caption{Matrices giving contributions near the left boundary.  A `$*$' represents a nonzero entry.}
\label{leftboundary}
\end{table}

\begin{table}
\begin{tabular}{ccccc}
$a=0$:
  & $\left(\begin{array}{cccc} *&0&*&0 \\ 0&*&0&* \\ *&0&*&0 \\ 0&*&0&* \end{array}\right)$ 
  & $\left(\begin{array}{cccc} 0&0&*&0 \\ 0&0&0&0 \\ 0&0&*&0 \\ 0&0&0&0 \end{array}\right)$ 
  & $\left(\begin{array}{cccc} 0&0&0&0 \\ *&*&*&* \\ 0&0&0&0 \\ *&0&0&* \end{array}\right)$ 
  & $\left(\begin{array}{cccc} 0&0&0&0 \\ 0&0&*&0 \\ 0&0&0&0 \\ 0&0&0&0 \end{array}\right)$ \\[27pt]
  & ${M}_{a,b}^{-1}{M}_{a,b+2}$ & ${M}_{a,b}^{-1}{M}_{a,b+6}$ 
     & ${M}_{a,b}^{-1}{M}_{a-1,b}$  
     & ${M}_{a,b}^{-1}{M}_{a-1,b+2}$ \\
  & ${M}_{a,b}^{-1}{M}_{a,b+4}$ & & ${M}_{a,b}^{-1}{M}_{a-2,b}$
     & ${M}_{a,b}^{-1}{M}_{a-2,b+2}$\\[20pt]
$a=1$:
  & $\left(\begin{array}{cccc} *&0&0&0 \\ *&0&*&0 \\ *&0&*&0 \\ *&0&0&0 \end{array}\right)$ 
  & $\left(\begin{array}{cccc} 0&0&0&0 \\ 0&0&*&0 \\ 0&0&*&0 \\ 0&0&0&0 \end{array}\right)$ 
  & $\left(\begin{array}{cccc} 0&0&0&0 \\ 0&0&0&0 \\ 0&0&0&0 \\ 0&0&0&0 \end{array}\right)$  
  & \\[27pt]
  & ${M}_{a,b}^{-1}{M}_{a-1,b}$ & ${M}_{a,b}^{-1}{M}_{a-1,b+2}$ 
     & ${M}_{a,b}^{-1}{M}_{a-2,b}$ &  \\
  & & & ${M}_{a,b}^{-1}{M}_{a-2,b-2}$ & \\[20pt]
$a=2$:
  & $\left(\begin{array}{cccc} *&0&0&0 \\ *&0&*&0 \\ *&0&*&0 \\ *&0&0&0 \end{array}\right)$ 
  & $\left(\begin{array}{cccc} 0&0&0&0 \\ 0&0&*&0 \\ 0&0&*&0 \\ 0&0&0&0 \end{array}\right)$ 
  &   
  & \\[27pt]
  & ${M}_{a,b}^{-1}{M}_{a-2,b}$ & ${M}_{a,b}^{-1}{M}_{a-2,b+2}$ 
     &  &  \\
  & & &  & \\[20pt]
\end{tabular}
\caption{Matrices giving contributions near the top boundary.  A `$*$' represents a nonzero entry.}
\label{topboundary}
\end{table}

{\bf Example 1}: It is well known that the TTW system is quantum superintegrable in the case $k=2$, $(p=2,\ q=1)$, \cite{TTW, KMPog10}. The generating operators, expressed in Cartesian coordinates are 
$$H=\partial ^2_x+\partial ^2_y+\alpha(x^2+y^2)+\beta \frac{(x^2+y^2)}{ (x^2-y^2)^2} +\gamma
\frac{(x^2+y^2)}{ 4x^2y^2}$$
$$-L_2=(x\partial _y-y\partial _x)^2+4\beta\frac{x^2y^2}{ (x^2-y^2)^2} + \gamma
\frac{(x^4+y^4)}{4 x^2y^2}+\beta+\frac{\gamma}{2},$$
$${\tilde L}=(\partial ^2_x-\partial ^2_y)^2+(2\alpha x^2 + 2\beta \frac{(x^2+y^2)}{ (x^2-y^2)^2} -
\gamma\frac {(x^2-y^2)}{2 x^2y^2})\partial ^2_x +$$
$$(-4\alpha xy +\frac {8\beta xy}{ (x^2-y^2)^2})\partial _x\partial _y + (2\alpha y^2 + 2\beta
\frac{(x^2+y^2)}{ (x^2-y^2)^2} + \gamma \frac {(x^2-y^2)}{2 x^2y^2})\partial ^2_y$$
$$+(2\alpha x - \frac{\gamma}{ x^3})\partial _x+(2\alpha y - 
\frac{\gamma }{ y^3})\partial _y+\alpha^2(x^2-y^2)^2 + \frac{\beta^2}{ (x^2-y^2)^2} + 
\frac{\gamma^2(x^2-y^2)}{16 x^4y^4}^2+$$
$$
8\alpha\beta \frac{x^2y^2}{ (x^2-y^2)^2} + \frac{\beta\gamma}{ 2x^2y^2} + 3\gamma(\frac{1}{2 x^4} + 
\frac{1}{2 y^4}).$$
By expressing the 4th order symmetry $\tilde L$ in polar coordinates and converting to canonical form we can read off the functions $A,B,C,D$  and then determine $F,\ G$ and the nonzero expansion coefficients. The results are
$$ {C}_{-2,1} = \left( \begin{array}{c} -44-4L_2 \\ 4(\gamma-\beta) \\ 0 \\ -28-8L_2 \end{array} \right),\    {C}_{-1,1} = \left( \begin{array}{c} -2H \\ 0 \\ 0 \\ -4H \end{array} \right),\ {C}_{0,1} = \left( \begin{array}{c} 0 \\ 0\\ 0\\ 2\alpha \end{array} \right).$$
It is easy to check that these terms satisfy all our recurrence relations.

{\bf Example 2}: The
nonzero vectors in the solution to the recurrence for the $k=1/3$ quantum TTW system are $C_{-1,1}$, $C_{-1,3}$, $C_{0,3}$ and $C_{0,1}$.
The solution to the recurrence depends linearly on two parameters that can be taken to be
$A_{-2,1,0}$  and $B_{-2,0,1}$.  To obtain solutions for $A$, $B$, $C$ and $D$ that are polynomial in
$H$ and $L_2$, we must choose the free parameters so as to cancel any demoninators.  With the choice
\[
B_{-2, 0, 1} = 36H(18L_2+13) \quad \mbox{and} \quad
A_{-2, 1, 0} = 8(81{L_2}^2+765L_2+274),
\]
we find that $\tilde L$ is a 6th order symmetry and the expressions for $A$, $B$, $C$ and $D$ 
are given below written in terms of the $u_1=R$ and $u_2=\theta$ coordinates.

\begin{eqnarray*}
 A &=& 8  \left( 81 {{L_2}}^{2}+765 {L_2}+274 \right) {{e}^{-
2 {u_1}}}\sin^3 \left( \frac23 {u_2} \right) 
-144  {(\gamma-\beta)}  \left( 9 {L_2}+26 \right) {{e}^{-2 {u_1}
}}\sin \left( \frac23 {u_2} \right) \cos \left( \frac23 {u_2} \right)
 \\
& & {} \
 +6
  \left( 6  \left( 23+9 {L_2} \right)  {(\gamma+\beta)}-81 { {(\gamma-\beta)}}^{
2}-81 {{L_2}}^{2}-765 {L_2}-274 \right) \sin \left( \frac23 {
u_2} \right) {{e}^{-2 {u_1}}}
\end{eqnarray*}
\begin{eqnarray*}
B &=& 
+40  \left( 8+81 {{ L_2}}^{2}+135 { L_2}
 \right) {{ e}^{-2 { u_1}}}   \sin^2 \left( \frac23{ u_2}
 \right)   \cos \left( \frac23{ u_2} \right) 
 \\
& & {} \
+48  \left( 101+144 { L_2} \right) { (\gamma-\beta)
} {{ e}^{-2 { u_1}}}   \sin^2 \left( \frac23{ u_2} \right) 
 \\
& & {} \
\left( 12  \left( 315 { L_2}+229 \right) { (\gamma+\beta)}-2754 {{ (\gamma-\beta)
}}^{2}-80-810 {{ L_2}}^{2}-1350 { L_2} \right) {{ e}^{-2 { u_1}}} \cos \left( \frac23
 { u_2} \right)
 \\
& & {} \
+12  \left( 162 { (\gamma-\beta)}
 { (\gamma+\beta)}- \left( 450 { L_2}+319 \right) { (\gamma-\beta)} \right) {
{ e}^{-2 { u_1}}}
 \end{eqnarray*}
\begin{eqnarray*}
C &=& -40  \left( 8+81 {{ L_2}
}^{2}+135 { L_2} \right) {{e}^{-2 { u_1}}}   \sin^3
 \left( \frac23 { u_2} \right) + 144  \left( 27 { L_2}+8 \right) { (\gamma-\beta)} \cos \left( \frac23 { 
u_2} \right) \sin \left( \frac23 { u_2} \right) {{e}^{-2 { u_1}}}
 \\
& & {} \
+6  \left( 6  \left( -27 { L_2}-5 \right) { (\gamma+\beta)}+81 {{ (\gamma-\beta)}
}^{2}+405 {{ L_2}}^{2}+40+675 { L_2} \right)  {{e}^{-2 { u_1}}}\sin \left( \frac23 
{ u_2} \right)
 \\
& & {} \
 -72 { H}  \left( 18 { L_2}+13
 \right)  \sin^3 \left( \frac23 { u_2} \right)   +1296 { H} {(\gamma-\beta)} \sin
 \left( \frac23 { u_2} \right) \cos \left( \frac23 { u_2} \right)
 \\
& & {} \
+54 { H}  \left( -6 { (\gamma+\beta)}+18 { L_2}+13 \right) \sin
 \left( \frac23 { u_2} \right)
\end{eqnarray*}
\begin{eqnarray*}
 D &=& - \left( 12 
 \left( 81 {{L_2}}^{2}+423 {L_2}+40 \right)  {(\gamma+\beta)}-162 
 \left( 9 {L_2}+8 \right) { {(\gamma-\beta)}}^{2} \right) \cos \left( \frac23
 {u_2} \right) {{e}^{-2 {u_1}}}
 \\
& & {} \
+12  \left( 9 {L_2}+
40 \right)  \left( 9 {L_2}+1 \right)  {(\gamma-\beta)} \cos \left( \frac43 {
u_2} \right) {{e}^{-2 {u_1}}}+2  \left( 81 {{L_2}}^{2
}+765 {L_2}+274 \right) {L_2} \cos \left( 2 {u_2}
 \right) {{e}^{-2 {u_1}}}
 \\
& & {} \
 \left( 36  \left( -27 {L_2}-23 \right)  {(\gamma-\beta)}  {(\gamma+\beta)}+162 
{ {(\gamma-\beta)}}^{3}+6  \left( 81 {{L_2}}^{2}+441 {L_2}+40
 \right)  {(\gamma-\beta)} \right) {{e}^{-2 {u_1}}}
 \\
& & {} \
+81  \left( -2  \left( 3 {L_2}
+7 \right)  {(\gamma+\beta)}+9 { {(\gamma-\beta)}}^{2} \right) {H} \cos \left( \frac23 {u_2} \right) +486  \left( {L_2}+2 \right) {H} {
(\gamma-\beta)} \cos \left( \frac43 {u_2} \right)
 \\
& & {} \
 + \left( 81 {{L_2}}^{2}+
441 {L_2}+40 \right) {H} \cos \left( 2 {u_2} \right)
\end{eqnarray*}

These give the operator $\hat L$ (\ref{generalLform}).  In order to construct the symmetry operator in standard
form (\ref{standardLform}), the $H$ and $L_2$, which have been treated as parameters throughout the calculation, must be moved to the
right.  For example, after expanding $C$, the coefficient of $HL_2$ is
\[ 
-1296 \sin^3 \left( \frac23 u_2 \right) + 972 \sin \left( \frac23 u_2 \right) 
\]
and so this contributes the term
\[ 
\left(-1296 \sin^3 \left( \frac23 u_2 \right) + 972 \sin \left( \frac23 u_2 \right) \right)
 \frac{\partial}{\partial u_2} H L_2
\]
to the differential operator $\tilde L$, in which $H$ and $L_2$ are now treated as differential operators.
We have used Maple to verify that the this operator does in fact commute with the $k=1/3$ quantum TTW Hamiltonian.

\section{Discussion}
Key to our method for proof of superintegrability is the canonical form for symmetry operators of all orders. It enables us to replace the computation of the commutator of $H$ with operators of arbitrary high order by verification of equations (\ref{eqn1}) and (\ref{eqn2}). In these equations $H$ and $L_2$ can be treated as parameters until the very last step when the canonical form is reinterpreted as an operator. Since (\ref{eqn1}) and (\ref{eqn2}) are linear and homogeneous  in $F$ and $G$ the solutions of these equations form a vector space. There are, of course, many solutions but most are not polynomials in $H, L_2$. To prove superintegrability we have to find a nontrivial  solution $F(u_1,u_2, H, L_2)$, $G(u_1,u_2,H, L_2)$ that has polynomial dependence on $H,L_2$. If there is one such solution, there will be an infinite number of others, since any polynomial function of a finite symmetry is a finite symmetry, as is the commutator of $L_2$ with a finite symmetry. To prove superintegrability we need find only one such solution. We choose the simplest ansatz that leads to success. The method we employ will not necessarily lead to the symmetry operator of lowest order. 

Our strategy is to postulate a set of basis functions  and to expand $F$ and $G$ in terms of them. The basis has to be chosen so that (\ref{eqn1}), (\ref{eqn2}) reduce to a set of recurrence relations between the coefficients of the basis functions. We will succeed if we can find some nonzero solution of these recurrences such that only a finite number of the coefficients are nonzero. The coefficients will then be rational functions of $H,\ L_2$, but arbitrary up to a scale factor $K(H,L)$. We choose $K$ such that all coefficients become polynomials in $H,L_2$, and then we are done! We have used two different methods to solve the recurrences in our two examples. The template method for the more complicated TTW problem is more general and its step-by-step evaluation of the expansion coefficients probably makes it the preferred tool  to treat additional examples.

How can one determine an appropriate set of basis functions? In the examples appearing in this paper we used the known expressions for the corresponding classical superintegrable systems as computed in \cite{KMPog10} to determine the basis for the classical constants of the motion, and then used the same basis for the quantum system. This worked although the classical expansion coefficients differed from the quantum coefficients, as would be expected.  It is clear that the methods of this paper will apply to many Hamiltonian systems, but each system will have its own peculiarities. Also, the canonical form for symmetry operators can clearly be extended to higher dimensions in the cases where the separable coordinates are of the subgroup type as treated in \cite{KKM10}. Of particular interest is  the relation between the classical constants of the motion and the quantum symmetries.  We intend to pursue these lines of inquiry.

\end{document}